\documentclass[iop]{emulateapj}
\usepackage{graphicx}
\usepackage{epstopdf}
\usepackage{natbib}
\usepackage{amsmath}

\providecommand{\e}[1]{\ensuremath{\times 10^{#1}}}
\usepackage{url}

\shortauthors{Davenport et al.}
\shorttitle{Short Period M Dwarf Binary}

\begin{document}

\title{The Very Short Period M Dwarf Binary SDSS J001641-000925}

\author{James R. A. Davenport\altaffilmark{1,2}}
\author{Andrew C. Becker\altaffilmark{2}}
\author{Andrew A. West\altaffilmark{3}}
\author{John J. Bochanski\altaffilmark{4}}
\author{Suzanne L. Hawley\altaffilmark{2}}
\author{Jon Holtzman\altaffilmark{5}}
\author{Heather C. Gunning\altaffilmark{2}}
\author{Eric J. Hilton\altaffilmark{6}}
\author{Ferah A. Munshi\altaffilmark{2}}
\author{Meagan Albright\altaffilmark{2}}

\altaffiltext{1}{Corresponding author: jrad@astro.washington.edu}
\altaffiltext{2}{Department of Astronomy, University of Washington, Box 351580, Seattle, WA 98195}
\altaffiltext{3}{Astronomy Department, Boston University, 725 Commonwealth Ave, Boston, MA 02215, USA}
\altaffiltext{4}{Astronomy and Astrophysics Department, Pennsylvania State University, 525 Davey Laboratory, University Park, PA 16802}
\altaffiltext{5}{Department of Astronomy, New Mexico State University, Box 30001, Las Cruces, NM 88003, USA}
\altaffiltext{6}{Dept. of Geology and Geophysics and Institute for Astronomy, University of Hawaii at Manoa, Honolulu, HI 96822}

\begin{abstract}
We present follow-up observations\footnote{Based on observations obtained with the Apache Point Observatory 3.5-meter telescope, which is owned and operated by the Astrophysical Research Consortium.
This paper includes data gathered with the 6.5 meter Magellan Telescopes located at Las Campanas Observatory, Chile. 
Support for the design and construction of the Magellan Echellette Spectrograph
was received from the Observatories of the Carnegie Institution of Washington,
the School of Science of the Massachusetts Institute of Technology, and the
National Science Foundation in the form of a collaborative Major Research
Instrument grant to Carnegie and MIT (AST0215989).
} and analysis of the recently discovered short period low-mass eclipsing binary, SDSS J001641-000925. 
With an orbital period of 0.19856 days, this system 
has one of the shortest known periods for an M dwarf binary system. 
Medium-resolution spectroscopy and multi-band photometry for the system are presented. Markov chain Monte Carlo modeling of the light curves and radial velocities yields estimated masses for the stars of 
M$_{1} =0.54\pm0.07$ M$_\odot$ and 
M$_2 = 0.34 \pm0.04$ M$_\odot$, and radii of 
R$_1 =0.68\pm.03$ R$_\odot$ and 
R$_2 = 0.58\pm0.03$ R$_\odot$ respectively. This solution places both components above the critical Roche overfill limit, providing strong evidence that SDSS J001641-000925 is the first
verified M-dwarf contact binary system.  Within the follow-up
spectroscopy we find signatures of non-solid body rotation velocities,
which we interpret as evidence for mass transfer or loss within the
system.  In addition, our photometry samples the system over 9 years,
and we find strong evidence for period decay at the rate of $\dot{P}\sim$8 s yr$^{-1}$.  Both of these signatures raise the intriguing
possibility that the system is in over-contact, and actively losing
angular momentum, likely through mass loss.  This places SDSS
J001641-000925 as not just the first M-dwarf over-contact binary, but
one of the few systems of any spectral type known to be actively
undergoing coalescence.  Further study SDSS J001641-000925 is on-going
to verify the nature of the system, which may prove to be a unique
astrophysical laboratory. 
\end{abstract}


\section{Introduction}
M dwarfs, with masses between $\sim$0.08 and 0.6 M$_\odot$, are the most numerous and faintest main sequence stars in the Galaxy. The high number density and very long main sequence lifetimes of M dwarfs allow them to be used as detailed tracers of the nearby galactic stellar population \citep{bochanski_gal,bochanski2010}. These stars are also well known for their high magnetic activity levels, characterized by frequent violent flaring outbursts \citep[e.g.][]{k09,davenport2012}, and H$\alpha$ emission \citep{west2008,westdr7,hilton10}.

Eclipsing binary systems provide us with the valuable ability to directly measure the masses and radii of the two component stars. As discussed in \cite{becker2011}, the sample of known binary systems composed of two M dwarfs is very small. Their intrinsically low luminosity and small radii conspire to make M dwarf binaries difficult to detect in variability surveys. The fraction of M dwarfs in binary systems is also much lower than for higher mass stars. \citet{2006ApJ...640L..63L} has estimated that as few as $\sim$10\% of M dwarfs in the Galactic field may be in binary (or higher number) configurations.

The M dwarf binary and single-star systems that are well studied indicate that stellar structure models do not accurately predict the observed radii \citep{ribas2006,morales2010}. Radii are often observed to be $\sim$10--20\% larger than predicted for a given mass. This may be due to limitations of the stellar models used, or to a fundamental difference in the interior structure of M dwarfs. Magnetic activity is frequently identified as a potential culprit, causing the radii to be enlarged due to the inhibition of convection or decreased heat flux  from magnetic spots \citep[e.g.][]{chabrier2007,bochanski2011}. Long period M dwarf binaries ($P\gtrsim10$ days) are ideal systems for radii determinations, as they are not influenced by spin--orbit coupling or tidal distortion \citep{mazeh2008}. The above discrepancies will only be resolved by the addition of many accurately measured radii to the known sample over a wide range in mass and period.

There also appears to be an absolute lower-limit on the orbital period for contact binary star systems, though no consensus about its origin has been reached. This short period limit at $\sim$0.22 days may be due to both components being fully convective \citep{rucinski1992}. At shorter periods, these stars cannot adhere to their mass-radius relationship while simultaneously staying within the Roche equipotential geometry. However, as \citet{rucinski1992} points out, this does not appear to be the primary effect at work in creating the sharp 0.22 day period limit. Alternatively, \citet{jiang2012} have shown that unstable mass transfer for low mass contact binaries may result in rapid coalescence. The observed period limit would thus be a result of the very short lifetimes for such systems. Models of angular momentum loss in binary systems, however, indicate that even given favorable mass ratios and initial orbital periods, 6-13 Gyr timescales are typically required for low mass binaries to obtain periods as short as 0.22 days \citep{stepien2006b}. The observed period limit would then be a consequence of the finite age of the binary population.

Until recently, the shortest known M dwarf binary system period in the literature was BW3 V38 \citep{maceroni1997,maceroni2004}, a detached binary composed of two main sequence M3 dwarfs with an orbital period of 0.1984 days.  \cite{rucinski2008} found a similarly short period system, GSC 01387--00475, with a period of 0.2178 days, but composed of higher mass K3 and K5 stars and in a contact configuration. 

\begin{figure}[t]
\centering
\includegraphics[width=3.5in]{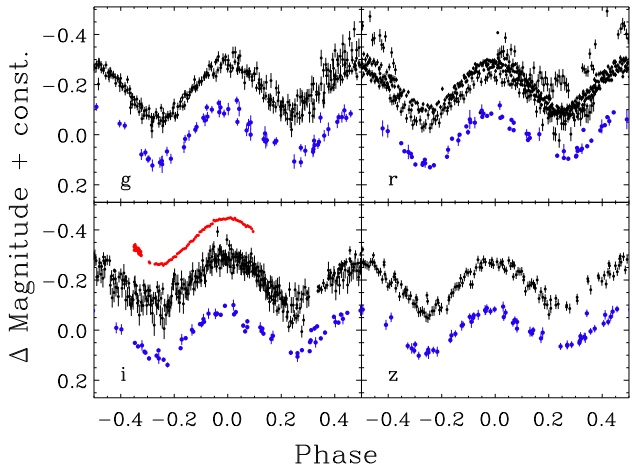}
\caption{Phased light curves in the five SDSS filter bandpasses for SDSS J001641-000925. Photometric errors are shown for all data. Blue points are from the Stripe 82 catalog, black are follow-up data from the NMSU 1-m in $griz$ bands. Red data in the $i$-band were obtained with the Agile camera at the APO 3.5-m.}
\label{lc}
\end{figure}

\cite{dimitrov2010} reported the shortest period M dwarf binary yet characterized, GSC 2314--0530 with a 0.192636 day period, and stellar masses of 0.51 M$_\odot$ and 0.26 M$_\odot$. This system has an orbital period only $\sim$8 minutes shorter than that of our target, SDSS J001641-000925. \cite{nefs2012} have recently discovered four binary systems with M dwarf colors that have periods significantly below the $\sim0.22$ day period limit, providing a major challenge to our understanding of the formation and lifetimes of such systems. 
The rarity of short period M dwarf binaries, and the lack of known contact M dwarf binary systems, indicates that they are difficult to produce, are so unstable that they are rapidly destroyed, or are still very difficult to detect.

A recent study of photometric variability by \citet[][hereafter B11]{becker2011} using the Sloan Digital Sky Survey \citep[SDSS,][]{york2000} revealed several new candidate M dwarf -- M dwarf eclipsing binary systems. 
B11 analyzed $\sim$4.8 million light curves and found 207 periodically variable objects, and a sharp drop in the number of systems with periods less than 0.22 days. Of these, two candidate eclipsing binary systems with sinusoidal light curves and  M dwarf $ugriz$ colors were found to have very short periods:  SDSS J200011.19+003806.5 at 0.14552 days, and SDSS J001641.03--000925.2 at 0.198561 days. Both of these objects warranted spectroscopic and photometric follow up to determine their fundamental parameters, and to confirm their status as short period eclipsing binaries.

In this paper we describe our detailed photometric and spectroscopic investigation of the brighter of these two objects, the 0.198561 day periodic variable SDSS J001641-000925. In \S2 we describe the photometric and spectroscopic data used in this study. We model basic parameters in \S3, and analyze the peculiarities of the binary system in \S4. We discuss the context of this over-contact binary in \S5.

\section{Observations}

SDSS J001641-000925 was discovered as a periodic variable object by B11 using time-domain data from the SDSS ``Stripe 82'' region \citep{ivezic2007,sesar2007}. This unresolved binary system has an SDSS ObjID=587731186205720773, and SpecObjID=314914764310446080. A period of 0.198561 days was determined from 61 epochs of $griz$ photometry, spanning more than 8 years of observations, using the SuperSmoother algorithm \citep{supersmoother,reimann1994}. This algorithm does not assume a light curve shape, and instead works to find the period that minimizes the phase folded light curve dispersion.
SDSS J001641-000925 has a median $r$-band magnitude of 16.39. A photometric spectral type of M1 was found using the median $riz$-bands and the covariance matrix technique from \cite{k09}, while a type of M0 was assigned using the ``Hammer'' spectral-typing analysis package \citep{covey2007}, which compares spectral data to a set of standard templates. The SDSS spectrum did not show any clear signatures of binarity, such as spectral line separation or large deviation from the M0 spectral template \citep{bochanski_templates}. In the following section we describe follow-up data obtained to characterize this short period object.


\subsection{Photometry}
Figure \ref{lc} contains all of the photometry for this system. In addition to the 61 epochs of SDSS Stripe 82 photometry, we obtained follow-up $i$-band photometry from the Agile high speed imaging camera on the Astrophysical Research Consortium  3.5-m telescope at Apache Point Observatory (APO), and $griz$-band photometry from the New Mexico State University (NMSU) 1-m. 

The SDSS photometry comes from the ``Stripe 82'' region, a narrow strip located along the celestial equator ($20^{\rm h} 24^{\rm m}< \alpha_{\rm J2000}<04^{\rm h} 08^{\rm m}$, $|\delta_{\rm J2000}|<1.266^\circ$) that has been repeatedly observed throughout the course of the SDSS. This catalog produced light curves for almost 8 million point sources with between $\sim$20 and $\sim$80 epochs. These data have been used for a variety of investigations, including studies of supernovae \citep{frieman2008}, stellar flares \citep{k09}, and characterizing multiwavelength point source variability \citep{sesar2007}. SDSS J001641-000925 was observed 61 times in all five photometric bands.

The Agile camera observed the system for $\sim$45\% of the orbital period (2.2 hours) on 2008 October 30, with a cadence of 90 seconds, and a median photometric error of 0.004 magnitudes. The $\sim$0.2 magnitude sinusoidal variation in the Agile light curve appeared in phase with the sparsely sampled Stripe 82 data (see Figure \ref{lc}).

Multi-band follow-up was provided by the automated NMSU 1-m telescope \citep{holtzman2010}. 
These $griz$ data were taken over a series of 16 half-nights in 2010 from November 13 to December 2. The observing strategy and exposure times varied between the 16 nights. During two nights we imaged the binary with consecutive $griz$ filter exposures. Single filters were used for each of the remaining 14 nights, broken down as follows: two nights of $g$, seven nights of $r$, three nights of $i$, and two nights of $z$-band. These phased light curves matched the well defined sinusoidal trends seen in the $griz$-bands from the SDSS data, with no major deviations e.g. from flares or star spots.


\begin{figure}[t]
\centering
\includegraphics[width=3.5in]{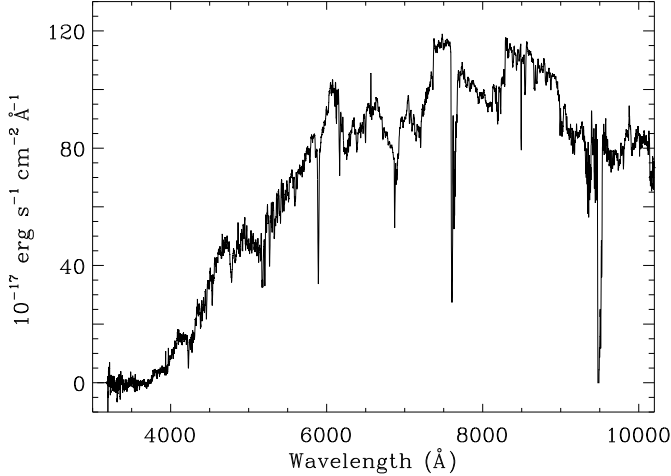}
\caption{One of the nine MagE spectra for SDSS J001641.03--000925.2, taken at a phase of $\phi=0.236$. The spectrum has been smoothed using a boxcar kernel of 7 pixels.}
\label{spec}
\end{figure}

\subsection{Spectroscopy}
SDSS spectra are the result of co-adding three to six individual frames, each with 9--15 minute exposure times \citep{york2000}. The time-domain SDSS spectra have previously been examined to search for emission line variability in M dwarfs \citep[e.g.][]{hilton10, kruse2010,bell2012}, and quasar variability \citep{vandenberk2004}. We obtained three time resolved spectra from 2002 October 12 using the SDSS Data Archive Server.\footnote{\url{http://das.sdss.org}} No spectral line splitting was apparent from the $R\sim1800$ resolution SDSS spectrum for this system. A single H$\alpha$ emission line was seen in each frame, but no radial velocity shift was resolved for any spectral feature.

Initial follow up spectroscopy was provided by the 6.5-m Magellan Clay telescope at Las Campanas Observatory, using the $R\sim4100$ resolution Magellan Echellette Spectrograph (MagE) on 2008 October 25. We obtained nine spectra with exposure times ranging from 600 to 900 seconds. These data were reduced using the MagE Spectral Extractor software package for IDL \citep[MASE;][]{bochanski2009}, and an example is shown in Figure \ref{spec}. The observations spanned $\sim$44\% of the orbital phase. 

\begin{figure}[t]
\centering
\includegraphics[width=3in]{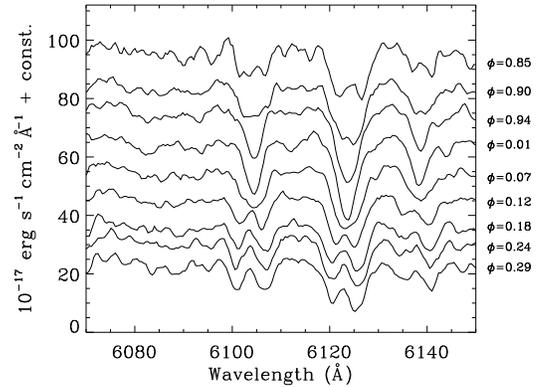}
\caption{Nine MagE spectra showing clear splitting in the Ca I lines at 6102\AA\ and 6122\AA. The MagE spectra were taken in order from top to bottom, and cover 44\% of the orbital phase, from $\phi=-0.15$ to $\phi=0.29$.}
\label{line}
\end{figure}

Additional spectroscopy was obtained with the APO 3.5-m Dual Imaging Spectrograph (DIS), using the R1200 grating with a resolution of $R\sim3500$, on 2011 October 20. These eight spectra had integration times of 900 seconds, and were timed to cover $\sim$50\% of the orbital phase that was not observed with MagE.


\section{Modeling the System}
Using the M dwarf spectral templates from \cite{bochanski_templates} we estimated the spectral type for both stars in this system. 
We scaled these template spectra, from types M0 to M3, by their approximate bolometric luminosities \citep{nlds2}, and created a normalized binary template spectrum for each of the 16 possible combinations of stars (M0+M0, M0+M1, ... M3+M3).  The average best fit template combination for the DIS and MagE data was M0+M1, which we adopted to seed the initial binary model.

\begin{figure}[t]
\centering
\includegraphics[width=3.5in]{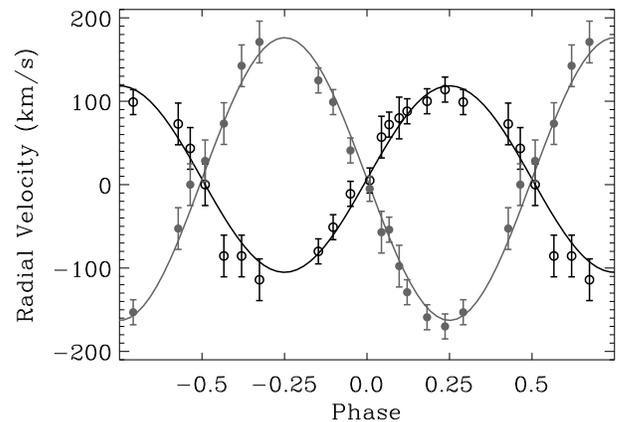}
\caption{Radial velocities as a function of orbital phase for both components of this system for the MagE and DIS data. The velocities for the primary star are open black circles, and the secondary star are solid grey circles. The best-fit model, as generated by the PHOEBE software, is overlaid. }
\label{rv_model}
\end{figure}

Velocities were measured for both stellar components by cross correlating the active M0 and M1 spectral templates from \cite{bochanski_templates} to each of the MagE and DIS spectra using a two-dimensional cross correlation code employing the TODCOR algorithm \citep{todcor}. Figure \ref{line} presents three absorption lines in the MagE data that show clear radial velocity splitting, indicative of a binary system. Velocities were also extracted from the spectra using the IDL broadening function code from \citet{rucinski2002}, and were consistent with the cross correlation results within the $\sim$15 km s$^{-1}$ TODCOR uncertainties. The radial velocity curve for the MagE and DIS data is shown in Figure \ref{rv_model} as a function of orbital phase, with the best-fit binary model (described below) overlaid.

\subsection{Initial Binary Model}

\begin{figure}[]
\centering
\includegraphics[width=3in]{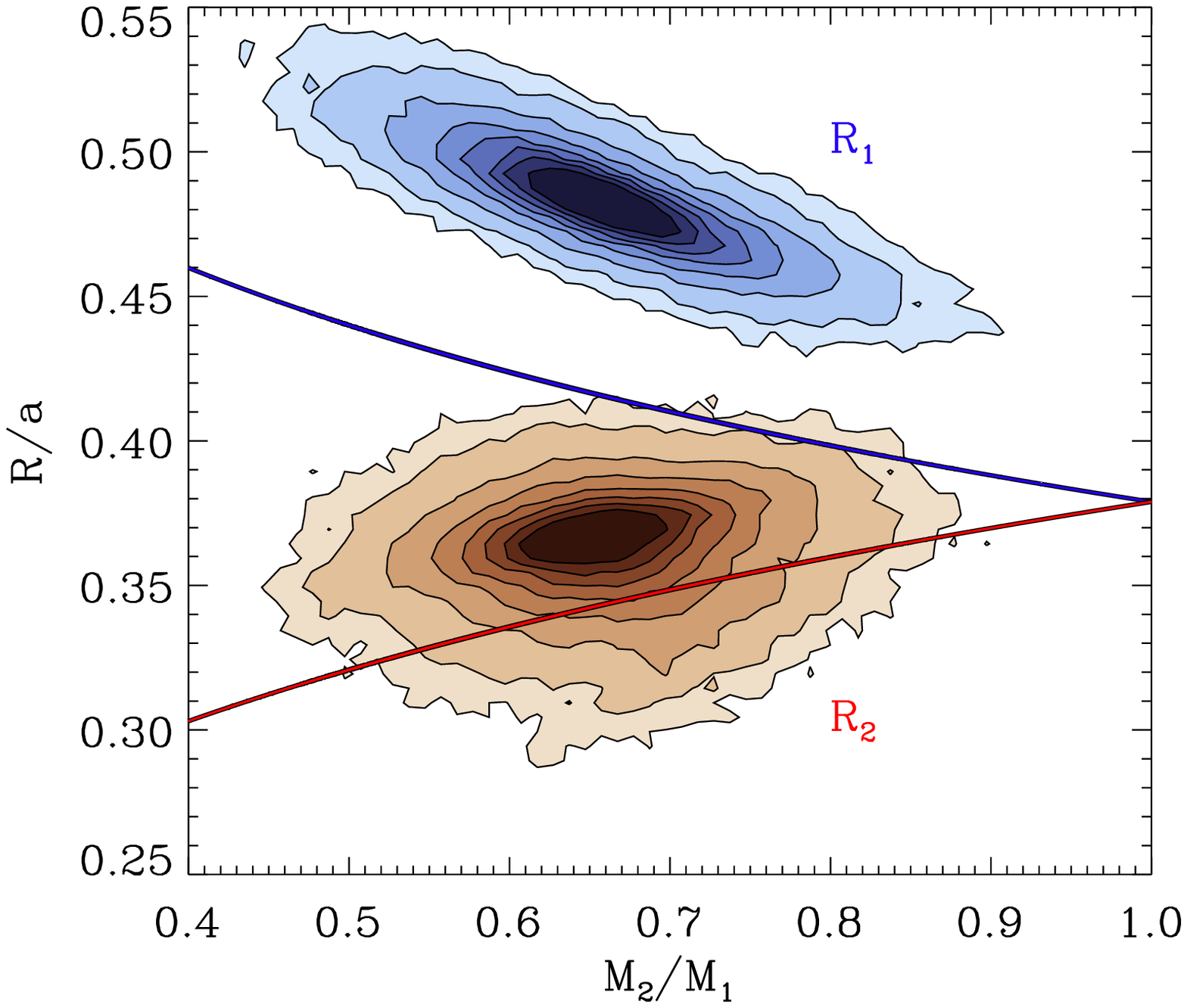}
\includegraphics[width=2.5in]{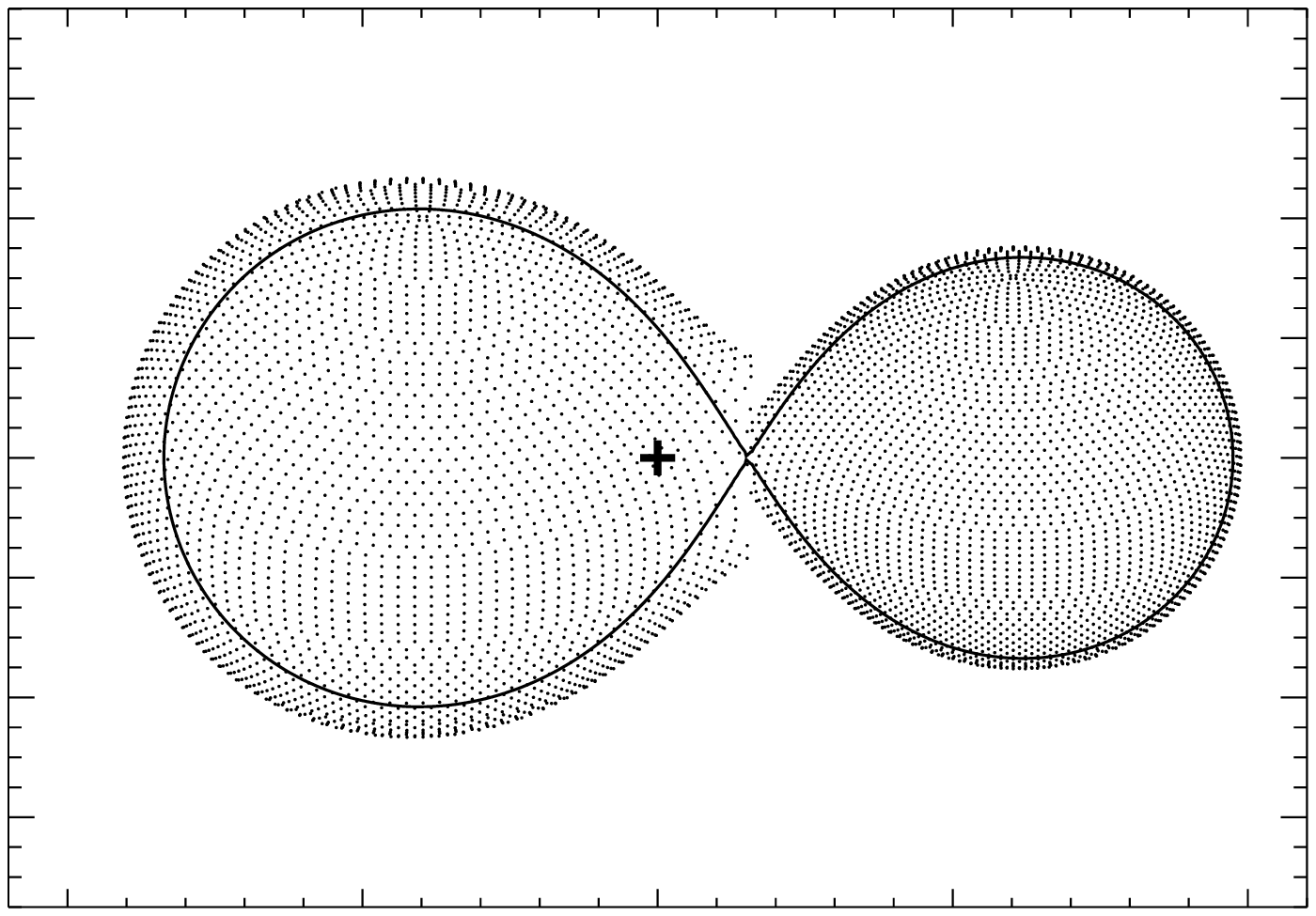}
\caption{Top: Probability distributions of the radii normalized by the semi-major axis, as a function of the mass ratio, for the primary star (blue contours) and secondary star (red contours) explored by the MCMC sampler using the {\tt Unconstrained} mode in PHOEBE. The solid blue and red lines show the effective Roche lobe radii from \citet{eggleton1983} for the primary and secondary stars, respectively. The effective radii for both stars are larger than the critical Roche lobe radii, indicating an over-contact binary scenario. 
Bottom: A three dimensional model for SDSS J001641-000925 at a phase of $\phi=0.25$ from the PHOEBE software using our best-fit values, with the primary star (left),  secondary star (right),  center of mass (cross), and the critical Roche equipotential (solid line).}
\label{mesh}
\end{figure}

In order to combine the many sources of data, we used the latest stable build of the PHOEBE\footnote{\url{http://phoebe.fiz.uni-lj.si/}} modeling software \citep[v0.32;][]{phoebe}. This implementation of the \cite{wilson1971} algorithm is an industry standard in modeling binaries, and is actively being developed and improved for use in modern binary studies \citep[e.g.][]{prsa2011} from missions such as the Kepler space telescope \citep{borucki2010}.

Due to the very short orbital period of the system we initially used the {\tt``Undetermined''} binary configuration in the PHOEBE software, which allowed solutions wherein both stars could potentially overflow their Roche lobes. We estimated the initial system parameters assuming a M0+M1 binary pair, as found by our spectroscopic fit described above, with masses of 0.6 and 0.5M$_\odot$, temperatures of 3800 and 3600 K, and an orbital separation of 1.4 R$_\odot$.

The binary parameters were converged manually using the ``Differential Corrections'' solution mode. We started by fitting only the photometry, assuming 70 degrees inclination and a circular orbit. The light curves for the $griz$ bands were fit simultaneously, and the effective temperatures and luminosities of the stars were estimated. The radial velocities from the MagE and DIS data were then included in the model, and the solution recomputed. The orbital parameters were then allowed to vary, and the solution was recomputed again using both the velocities and photometry as constraints.  Eccentricity remained fixed at 0. The orbital period from B11 for the system was adopted, with no time derivative or phase shift in the period. A logarithmic limb darkening profile, included with PHOEBE, was used. No surface spots were included in our binary solution.

\subsection{MCMC}

\begin{figure}[]
\centering
\includegraphics[width=3in]{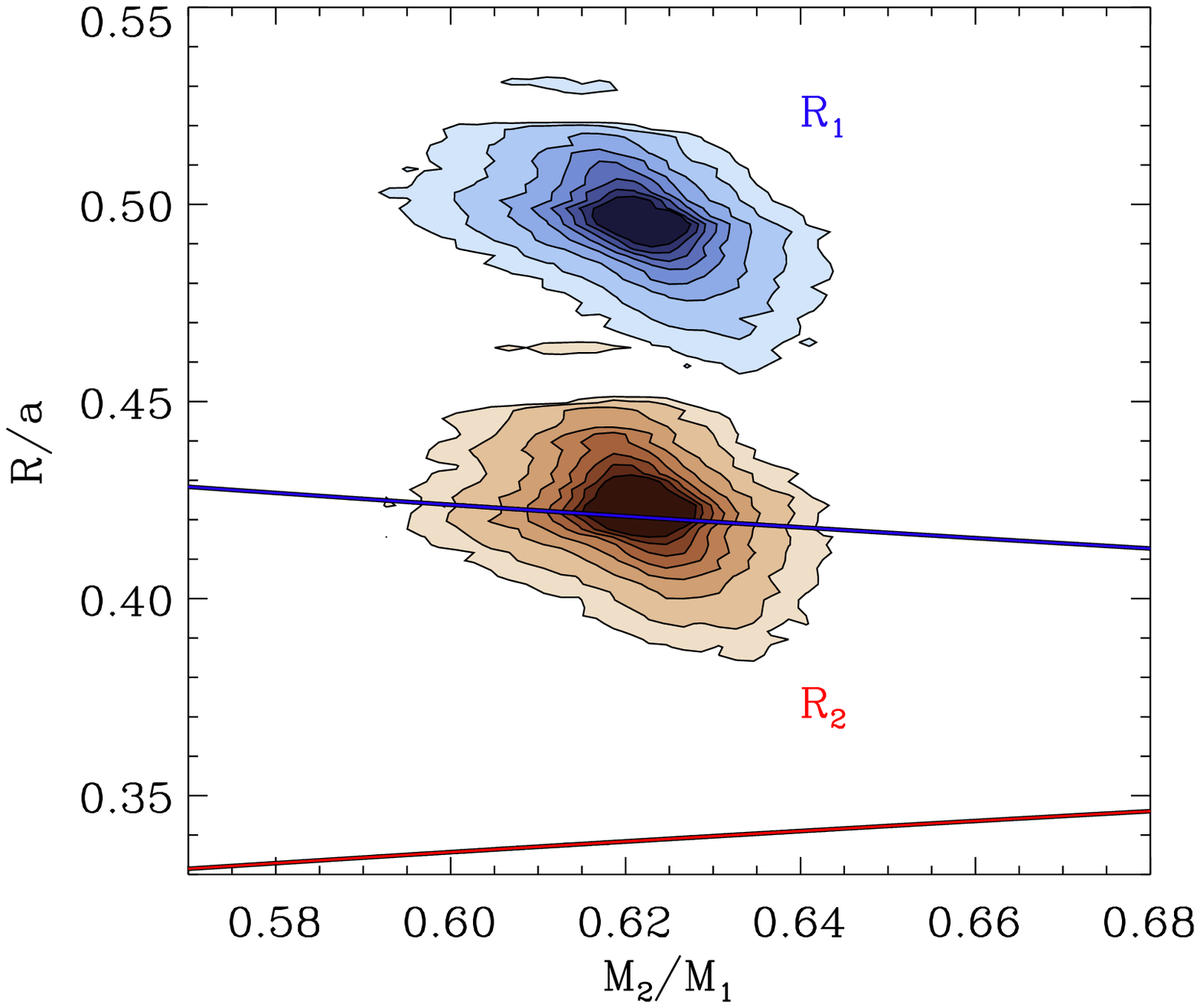}
\includegraphics[width=2.5in]{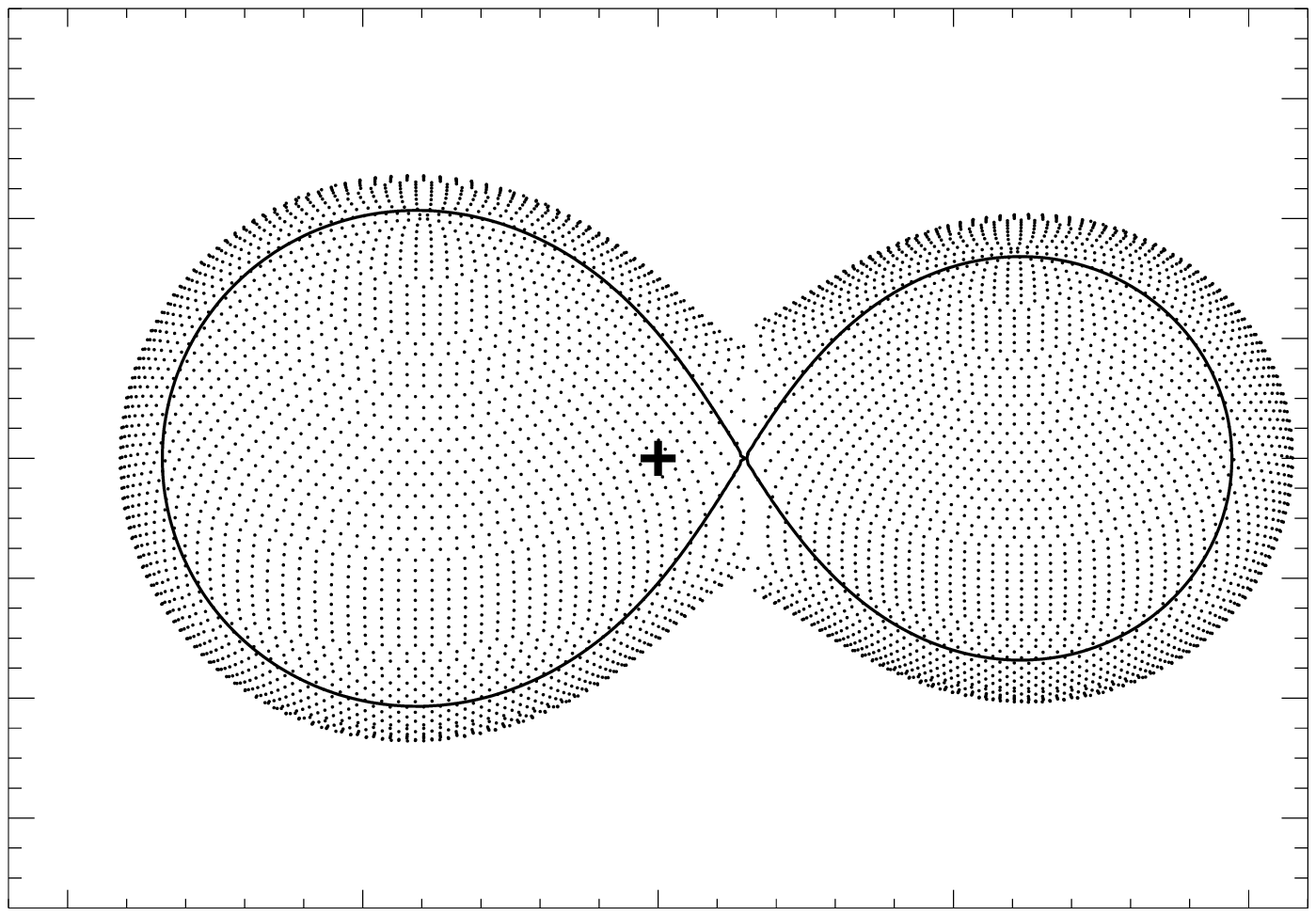}
\caption{Same as Figure \ref{mesh}, but for the W UMa {\tt Overcontact} mode in PHOEBE.}
\label{fillfac}
\end{figure}

\begin{figure*}[t]
\centering
\includegraphics[width=5in]{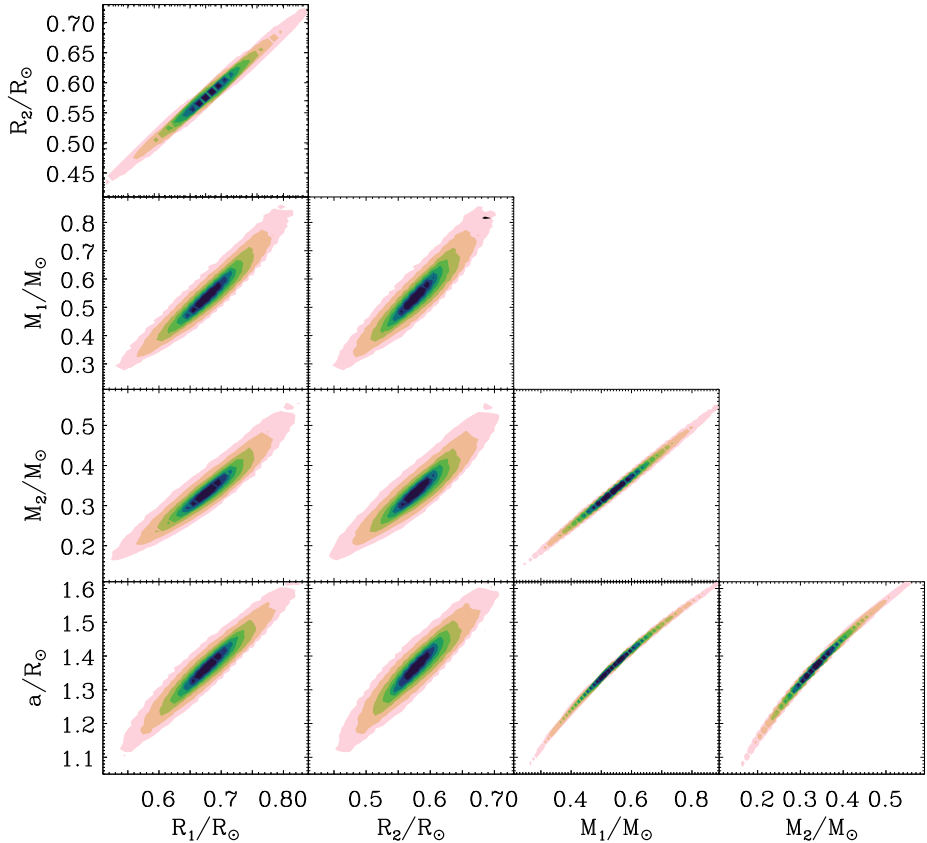}
\caption{Probability distributions for the masses, radii, and the semi-major axis for SDSS J001641-000925 as explored by our MCMC sampler using the W UMa {\tt Overcontact} mode in PHOEBE.}
\label{mcmc}
\end{figure*}

The result of our initial binary solution indicated that the system was in a contact configuration, with the primary star overflowing its Roche lobe. This would make SDSS J001641-000925 the first contact M dwarf binary system yet characterized. To determine the reliability of the contact solution, we employed a Markov Chain Monte Carlo (MCMC) parameter space search method.

We used the Python-based MCMC sampler {\it emcee} \citep{emcee} to adjust the free parameters $T_1$, $T_2$, $a$, $i$, $v_{sys}$, and the potentials for the two stars $\Omega_1$ and $\Omega_2$ \citep{kopal1959}.
Initial uncertainties for our MCMC runs were determined using the error prescription included in the PHOEBE v0.32 user manual for our initial binary model parameters. We then used the command line PHOEBE ``Scripter'' to compute the $\chi^2$, masses, and radii for each iteration (step). 

A total of 16 parameter space ``Walkers'' were used, and multi-threading with seven computer cores was implemented using the Python multiprocessing module. The first 1,000 steps were discarded, to allow the sampler an initial exploration of the parameter space. The MCMC was run for an additional 20,000 steps for each Walker, for a total of 320,000 iterations. 

We ran our MCMC search using both the ``{\tt Overcontact Binary of the W UMa type}'' mode and ``{\tt Unconstrained}'' mode in PHOEBE, as the constraints for each mode can affect the resulting parameter estimates, especially of the radii \citep[e.g.][]{bonanos2009,garcia2010}. 
The best-fit results for the free parameters in both modes were the same within the uncertainties. The secondary star radius, however, was very different between the modes. In Figure \ref{mesh} we show the radii versus mass ratio probability distributions, and a three dimensional illustration of the system at a phase of $\phi=0.25$, with the {\tt Unconstrained} mode MCMC runs. Figure \ref{fillfac} shows the MCMC results using the W UMa {\tt Overcontact} mode.  We note that the {\tt Overcontact} model \citep{wilson1971,ibvs5076} can converge on a physically incorrect solution \citep[e.g. see][]{pribulla2008}, particularly for systems with mass transfer or mass loss as discussed in \S4.2. Both modes, however, indicated that the system is in contact \citep{ibvs4460,ibvs5076}, and we present results for the {\tt Overcontact} W UMa mode as our preferred final model. The stars were found to have masses of M$_{1} =0.54\pm0.07$ M$_\odot$ and 
M$_2 = 0.34 \pm0.04$ M$_\odot$, which correspond to spectral types M1 and M3 \citep{nlds2},
and radii of 
R$_1 =0.68\pm.03$ R$_\odot$ and 
R$_2 = 0.58\pm0.03$ R$_\odot$. A grid of the resulting probability distributions for the masses, radii, and semi-major axis is shown in Figure \ref{mcmc}. 

We computed the correlation lengths for each of our free parameters using the autocorrelation function. The effective chain length is defined as the actual length of the MCMC chain after the 1,000 step ``burn-in'', divided by the correlation length, and must be $>>1$ for meaningful results \citep{tegmark2004}. The shortest effective chain length in our study was for the inclination parameter, with a value of 7,106. All other free parameters had effective chain lengths over 9,000, indicating that our parameter space sampling reliably converged on the correct solution, and that the resulting errors were realistic. We also tested the convergence of our MCMC runs with the Gelman-Rubin $\hat{R}$ statistic, as detailed in \S3.3 of \cite{ford2006}. All parameters converged to $\hat{R}<1.001$, indicating our MCMC effectively sampled the parameter space.

\begin{figure}[!t]
\centering
\includegraphics[width=3.5in]{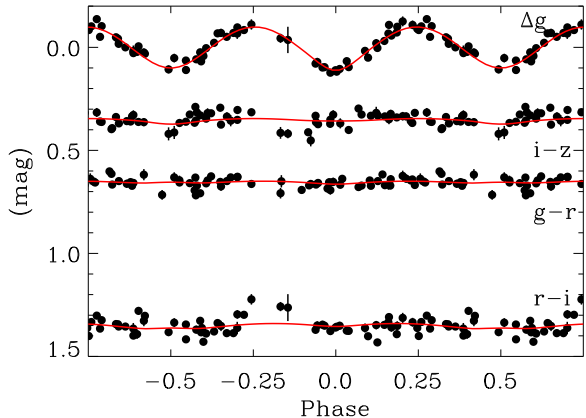}
\caption{The differential SDSS $g$-band, $(g-r)$, and $(r-i)$ colors as a function of orbital phase, along with the PHOEBE model fit (red).}
\label{lc_model}
\end{figure}

The final radial velocity solution is shown in Figure \ref{rv_model}, and the SDSS phase curves computed from the final PHOEBE model are shown in Figure \ref{lc_model}. The median model values for the system parameters are given in Table \ref{datatable}. Errors for each parameter were determined using the standard deviation of the probability distributions.

\section{Results}

\begin{deluxetable}{ccc}
\tablecolumns{3}
\tablecaption{The system parameters and uncertainties for SDSS J001641-000925 determined by our MCMC runs.}
\tablehead{
	\colhead{Quantity}&
	\colhead{Value} &
	\colhead{Uncertainty}
	}
\startdata
$P$ (day) & 0.198561 & 1\e{-6} \\
$t_0$ (day) & 0.002405 & 1\e{-6} \\
M$_1$ (M$_\odot$)  &  0.54 & 0.07 \\
M$_2$ (M$_\odot$)  &  0.34 & 0.04 \\
R$_1$ (R$_\odot$)  &  0.68 & 0.03 \\
R$_2$ (R$_\odot$)  &  0.58 & 0.03 \\
T$_{\rm eff,1}$ (K) & 4342. & 475. \\
T$_{\rm eff,2}$ (K) & 3889. & 579. \\
a (R$_\odot$)  &   1.37 & 0.06 \\
q (M$_2$/M$_1$)& 0.62 & 0.01 \\
$i$ (deg) & 53.3 &  1.1 \\
v$_{sys}$ (km s$^{-1}$) & -5.6 &  3.9
\enddata
\label{datatable}
\end{deluxetable}

\subsection{Binary System Parameters}
The illustration of our binary system in Figure \ref{fillfac} clearly indicates that both stars are filling their Roche lobes. We also show the probability distributions for the normalized radii of both stars. Overlaid is the critical Roche radius for both the primary and secondary stars, calculated using the \citet{eggleton1983} approximation. The MCMC probability distribution for the W UMa model mode showed both stars to be larger than their critical Roche radius with 100\% confidence. The primary star was 18\% larger than the critical Roche radius, and the secondary star 25\% larger. This raises the intriguing possibility that the system is in over-contact \citep{ibvs4460}, in which case we would expect to see non-solid body velocities from outflowing material leaving the system via the L2 point. This would be a very efficient method of angular momentum loss for the binary, and could lead to rapid decay of its orbit.

\begin{figure}[!t]
\centering
\includegraphics[width=3.5in]{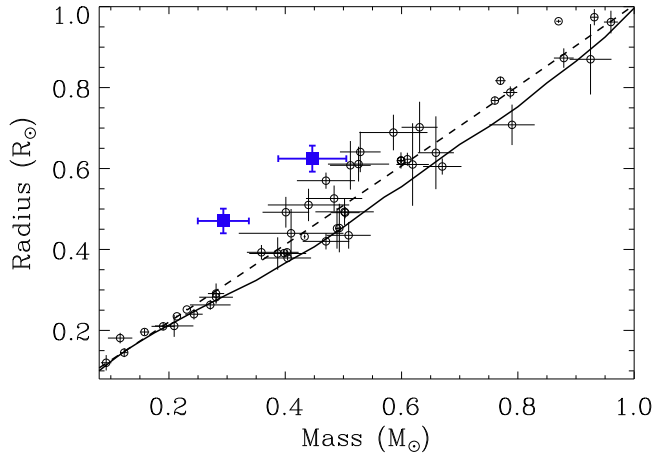}
\caption{The stellar radius as a function of mass for the compilation of single and binary stars from \cite{lopez2007}, as well as the system from \cite{becker2008}, in open black circles. Our PHOEBE solution for SDSS J001641-000925 is shown as filled blue squares. The solid black line is the \citet{baraffe1998} stellar evolution model, and the dashed line is the empirical fit from \citet{bayless2006}.}
\label{massradius}
\end{figure}

The masses and radii of SDSS J001641-000925 with respect to other low-mass eclipsing binaries are plotted in Figure \ref{massradius}. \cite{bayless2006} provided an empirical fit to the Mass--Radius relation using many low-mass binary systems with larger than predicted radii. 
Our system is much larger than the \cite{bayless2006} prediction at this mass range, due to the tidal distortion effects of the over-contact solution.

\subsection{H$\alpha$ Line Emission}

H$\alpha$ emission from the system was present in every observed spectrum. This is not an uncommon property of M dwarfs in general, whose magnetic activity is frequently characterized by strong and variable H$\alpha$ line emission \citep[e.g.][]{westdr7,bell2012}. However, the fraction of magnetically active field stars with spectral types M2 or earlier is less than 10\% \citep{westdr7}. The strength and frequency of this emission decreases as M dwarfs age, and the lifetime of magnetic activity for early type M dwarfs in the field is less than $\sim$2 Gyr \citep{west2008}. H$\alpha$ emission is also seen from M dwarfs in close binary systems with other dwarf stars \citep[e.g.][]{dimitrov2010} and with white dwarfs \citep{maxted1998,silvestri2006}. The higher fraction of active early type M dwarfs in close binaries as compared to the field is likely due to these stars rotating more rapidly than their isolated counterparts, inducing stronger and longer lived magnetic activity \citep{morgan2012,silvestri2006}.

\begin{figure}[!h]
\centering
\includegraphics[width=3.5in]{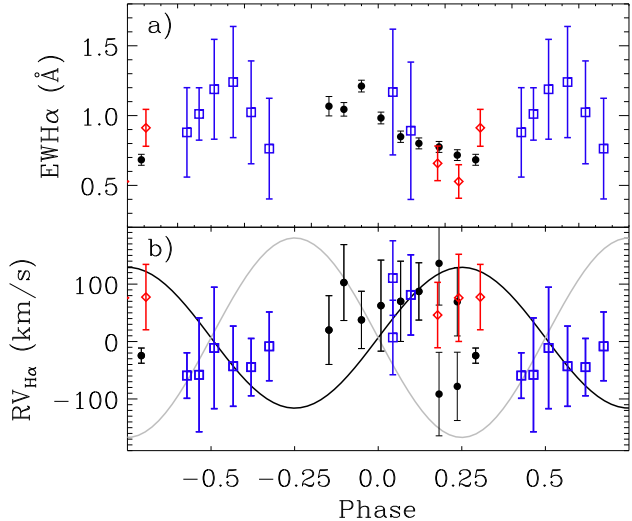}
\caption{a) The equivalent width of the H$\alpha$ line in each spectrum as a function of orbital phase for the MagE (filled black circles), DIS (open blue squares), and SDSS time-spectra (open red diamonds). b) The radial velocity of the H$\alpha$ line for each spectrum, with the orbital radial velocity solution from Figure \ref{rv_model} overlaid for comparison, using the same color coding and symbols as above.}
\label{ewha}
\end{figure}

We measured the H$\alpha$ equivalent width (EW) for each DIS, MagE, and SDSS spectra,  shown in Figure \ref{ewha}a as a function of orbital phase. The strength of emission is nearly constant at $\sim$1\AA, though a weak correlation in the line strength was found with orbital phase, with a higher H$\alpha$ EW at conjuncture ($\phi=0$ and $\phi=0.5$).

The H$\alpha$ line was observed to be broader than other features (e.g. the Ca I absorption lines in Figure \ref{line}), and clear splitting was only seen in a few spectra. The feature did not contribute significantly to the cross correlation in \S3, which was dominated by absorption features such as those shown in Figure \ref{line}. We measured the radial velocity of the H$\alpha$ line, shown in Figure \ref{ewha}b, by fitting two Gaussian profiles to the emission feature. The H$\alpha$ line broadly followed the velocity trend of the more massive stellar component, possibly due to the primary dominating the total H$\alpha$ flux. Line splitting was seen near quadrature ($\phi=0.25$) in the higher resolution MagE spectra, and the H$\alpha$ velocity profiles are consistent with emission from both stellar components in this system.

No other evidence for accretion, flaring, or outflow in the form of nebular emission line features was observed.  The line splitting seen in the MagE data at $\phi=0.25$ indicates that both stars are magnetically active. We would conventionally assume the H$\alpha$ emission comes from many distributed active regions on the stellar surfaces. 
However, the broadened H$\alpha$ emission, lower amplitude radial velocities, and intermittent line splitting suggests non-solid body velocities, and may indicate mass transfer between the stars and mass loss, likely through the L2 point as a result of the over-contact configuration. The slight increases in H$\alpha$ EW at conjuncture are reminiscent of those from the massive over-contact binary system RY Scuti, which are attributed to extended H$\alpha$ emission from outflows \citep{grundstrom2007} that appear to have brighter EW during eclipse ($\phi=0$ and $\phi=0.5$) when the system flux is lower. Higher resolution spectroscopic monitoring throughout the orbital period is necessary to accurately determine the geometry of the H$\alpha$ emission regions, and the possible active region timescales.

\begin{figure}[!t]
\centering
\includegraphics[width=3.5in]{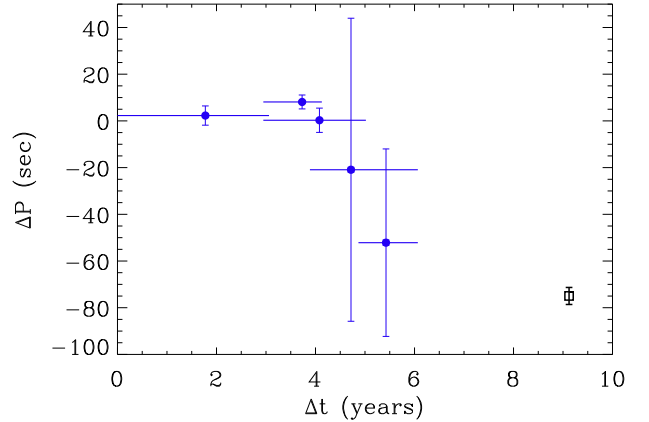}
\caption{The difference in orbital period between the B11 value and our measurements, using SDSS (blue) and NMSU (black) photometry, as a function of time. Periods shorter than the B11 value yield negative $\Delta P$. The first SDSS epoch, at $\Delta t =0$, had a MJD = 51075.316. Horizontal bars give the range of data for each SDSS subset, with each point placed at the mean epoch for the subset. Error bars were determined using a bootstrap method, described in the text.}
\label{pdot}
\end{figure}

\vspace{10mm}

\subsection{Period Decay}

The observed period of a close binary system should decrease over time, as the orbit decays and the system loses angular momentum from effects such as magnetic braking, tidal friction, mass transfer, mass loss, and in the most compact cases gravitational waves. Changes in the orbital period have been observed for several hundred W UMa-type binary systems. \citet{pilecki2007} measured period decreases for $\sim$30 short period systems, finding typical amplitudes of $\dot{P} \equiv \frac{dP}{dt} \sim -1$ sec year$^{-1}$. \cite{kubiak2006} searched the OGLE database \citep{ogle} for period changes from more than 500 short period binaries, finding a maximum amplitude of $\dot{P}=-0.4$ sec year$^{-1}$.

The SDSS Stripe 82 data were very sparsely sampled in time, and individual times of light curve minima were not observed. As such, we were unable to make a traditional Observed - Computed (O-C) diagram. Instead we broke the SDSS data into five subsets, spanning different ranges in time, each having at least 20 epochs.  The NMSU 1-m photometry were obtained in a single observing season, with many full orbits observed, and were used as a sixth subset.

To search for changes in the orbital period of SDSS J001641-000925, we measured the period for each time interval in our data with a phase dispersion minimization code, using the SuperSmoother algorithm \citep{supersmoother} as described in \citet{reimann1994}. We folded the each subset at periods ranging $\pm5$ minutes around the B11 value, in steps of 0.1 seconds.

In Figure \ref{pdot} we show the difference in period from the B11 value for each subset of $r$-band Stripe 82 and NMSU data. The time span for each Stripe 82 subset are shown as horizontal bars. A consistently decreasing period was found, with a total amplitude of $\sim$75 seconds, spanning the 9 years of observations, yielding $\dot{P} \sim -8$ sec year$^{-1}$. A similar approach was used to measure period decay for V1309 Sco in \citet{tylenda2011}.

We recomputed the period determination for each data subset using a bootstrap method, leaving one epoch out each time. Errors shown in Figure \ref{pdot} are the standard deviation of the bootstrap runs. The Stripe 82 period results are somewhat dependent on the time range chosen, as shown in Figure \ref{pdot}. The independent period measurement from the NMSU photometry, however, provides strong evidence of orbital decay for this system, though we note the caveat that this is dominated by the NMSU photometry.

\section{Discussion}
 
We have presented follow up spectroscopic and photometric observations of a low-mass binary, SDSS J001641-000925, whose orbital period is below the theoretical short period limit \citep{rucinski1992}. The components of this short period system have masses typical of spectral types of M1 and M3, and are in an over-contact configuration. Both stars have radii much larger than predicted by stellar isochrone models. These stars are the most ``oversized'' of any yet determined at this mass due to the over-contact configuration.

The strength of the H$\alpha$ line at $\sim$1\AA\ was not particularly high with respect to active stars of this spectral type. Higher H$\alpha$ flux is typically accompanied by strong magnetic activity and flaring for early type M dwarfs;  these are features that diminish as the stars age and lose angular momentum \citep{west2008}. The SDSS spectroscopic M dwarf sample, containing $\sim$70,000 spectroscopically confirmed M dwarfs \citep{westdr7} shows approximately half of stars with spectral types M0--M3 that are classified as ``active'' have EW(H$\alpha)\ge1.25$\AA. Thus the H$\alpha$ EW for this system is not usually high, nor does it appear particularly active. In fact, no signs of flaring were detected in more than 20 nights of observing. This is in strong contrast to the similar mass semi-detached binary system observed by \cite{dimitrov2010}, which showed six flares in a comparable amount of observing time. \citet{davenport2012} found that field stars in this spectral type range displayed photometric flares with amplitudes of $\Delta u\ge 1$ mag once every $\sim$2 days on average. 

However, the slight increase in H$\alpha$ EW during eclipse, as well as the broadened H$\alpha$ line profile with possibly non-solid body rotation velocities, together suggest that the emission feature may be the result of material being expelled from the outer Lagrange points. This would result in a complex H$\alpha$ line, with phase-dependent emission lines contributed from both stars, as well as possible P Cygni-like profiles from the outflowing material. Additional medium to high resolution spectroscopic monitoring will allow us to differentiate H$\alpha$ emission due to active surface regions from that of any nebular emission surrounding the system, as in RY Scuti \citep{grundstrom2007}.

Dynamical interactions between a binary system and other stars can decrease the binary orbital period through angular momentum transfer, or through three-body interactions \citep{kozai1962}. Indeed the incidence of contact binary systems is higher in stellar clusters than in the field \citep{1998AJ....116.2998R}, due to the more frequent three-body interactions. Recent models predict a timescale of 6-13 Gyr for an isolated low-mass binary system to undergo enough angular momentum loss to evolve from a stable orbital period of $\sim$1-2 days to below 0.22 days \citep{stepien2006a,stepien2006b}. Estimates of angular momentum loss from single stars, applied to short period binary systems, show that magnetic braking takes on the order of a Hubble time to compress these periods from 1--2 days to below the theoretical limit \citep[e.g.][]{maceroni2004}. With their discovery of several low mass systems below the predicted 0.22 day limit, \cite{nefs2012} demonstrated that initial orbital periods must have been $\sim$1 day in order for magnetic braking schemes to explain the short period systems. \cite{nefs2012} also show that the orbital evolutionary timescales and initial conditions for these rare short period binaries are still largely unconstrained, and that many possible mechanisms may be at work in forming such systems. Once the stars reach a contact scenario, unstable mass transfer via the filled Roche lobes, and mass loss through the outer Lagrange points, predicts a rapid evolution towards coalescence \citep{jiang2011}. This type of over-contact evolution \citep{ibvs4460} is thought to be a formation path for blue straggler stars in stellar clusters \citep{bradstreet1994,mateo1990}, and signatures include mass loss, changes in luminosity, and orbital decay.

Our measured period decay of approximately $|\dot{P}|\sim 8$ sec each year is larger than found for typical W UMa binary systems. The largest amplitude period decay measured for a contact binary is that of V1309 Sco, a merger system with an exponentially decreasing period with an amplitude of $|\dot{P}|\sim$300 sec year$^{-1}$ \citep{tylenda2011}. This system exhibited an extreme nova-like outburst, similar to V838 Mon, as it underwent a catastrophic merging event in 2008.  Assuming a linear period decay of $|\dot{P}|\sim8$ seconds per year, we would expect SDSS J001641-000925 to have a lifetime of $\sim$10$^3$ years. However, if the period evolution follows an exponential decay as in the higher mass V1309 Sco system, the binary merger timescale may be as short as $\sim10^2$ years. 

This amplitude of angular momentum loss required by the observed period decay can't be explained solely by magnetic braking, and the stars are too widely separated for gravitational waves to expel significant energy \citep{chau1978}. The angular momentum loss is likely due to  the system being in an over-contact configuration and having filled its critical Roche surface, with mass being expelled from the outer Lagrangian points. This would carry much more angular momentum from the system than magnetic breaking or winds alone, and would lead to a rapid coalescence into a single star \citep{li2004}. However, we cannot exclude the possibility of mass transfer from the primary to the secondary star with little to no loss of angular momentum to the system. This mass redistribution could instead lead to a binary evolution as described by \citet{stepien2006a}, with the orbital period increasing, and binary coalescence proceeding on a much more gradual nuclear or thermal timescale \citep{li2004}. Our follow up study of the period evolution will help discern between these two possibilities.
 
We emphasize the rarity of this system, as SDSS J001641-000925 is the first true over-contact M dwarf binary known. The critical question remaining for the system is whether it is evolving rapidly towards coalescence due to the over-contact configuration. Additional high precision follow-up photometry, as well as searching existing time domain surveys, is underway to conclusively determine whether the period is decaying at an exponential rate, as was the case for V1309 Sco before its dramatic merger.

\acknowledgements
The authors would like to thank Dr. John Wisniewski for fruitful discussions of low mass stars in binary systems, Dr. Adam Kowalski and Dr. Sarah Schmidt for assistance with observing, Mr John Ruan for help with Python, and the referee, Dr. Slavek Rucinski, for comments and suggestions that greatly improved this manuscript.
JRAD and ACB acknowledge support from NASA ADP grant NNX09AC77G. SLH and EJH acknowledge support from NSF grant AST 0807205.

This publication makes use of data products from the Two Micron All Sky Survey, which is a joint project of the University of Massachusetts and the Infrared Processing and Analysis Center/California Institute of Technology, funded by the National Aeronautics and Space Administration and the National Science Foundation.

Funding for SDSS-III has been provided by the Alfred P. Sloan Foundation, the Participating Institutions, the National Science Foundation, and the U.S. Department of Energy Office of Science. The SDSS-III web site is \url{http://www.sdss3.org/.} SDSS-III is managed by the Astrophysical Research Consortium for the Participating Institutions of the SDSS-III Collaboration including the University of Arizona, the Brazilian Participation Group, Brookhaven National Laboratory, University of Cambridge, University of Florida, the French Participation Group, the German Participation Group, the Instituto de Astrofisica de Canarias, the Michigan State/Notre Dame/JINA Participation Group, Johns Hopkins University, Lawrence Berkeley National Laboratory, Max Planck Institute for Astrophysics, New Mexico State University, New York University, Ohio State University, Pennsylvania State University, University of Portsmouth, Princeton University, the Spanish Participation Group, University of Tokyo, University of Utah, Vanderbilt University, University of Virginia, University of Washington, and Yale University.


\begin{thebibliography}{72}
\expandafter\ifx\csname natexlab\endcsname\relax\def\natexlab#1{#1}\fi

\bibitem[{{Baraffe} {et~al.}(1998){Baraffe}, {Chabrier}, {Allard}, \&
  {Hauschildt}}]{baraffe1998}
{Baraffe}, I., {Chabrier}, G., {Allard}, F., \& {Hauschildt}, P.~H. 1998, \aap,
  337, 403

\bibitem[{{Bayless} \& {Orosz}(2006)}]{bayless2006}
{Bayless}, A.~J., \& {Orosz}, J.~A. 2006, \apj, 651, 1155

\bibitem[{{Becker} {et~al.}(2008){Becker}, {Agol}, {Silvestri}, {Bochanski},
  {Laws}, {West}, {Basri}, {Belokurov}, {Bramich}, {Carpenter}, {Challis},
  {Covey}, {Cutri}, {Evans}, {Fellhauer}, {Garg}, {Gilmore}, {Hewett},
  {Plavchan}, {Schneider}, {Slesnick}, {Vidrih}, {Walkowicz}, \&
  {Zucker}}]{becker2008}
{Becker}, A.~C., {et~al.} 2008, \mnras, 386, 416

\bibitem[{{Becker} {et~al.}(2011){Becker}, {Bochanski}, {Hawley}, {Ivezi{\'c}},
  {Kowalski}, {Sesar}, \& {West}}]{becker2011}
{Becker}, A.~C., {Bochanski}, J.~J., {Hawley}, S.~L., {Ivezi{\'c}}, {\v Z}.,
  {Kowalski}, A.~F., {Sesar}, B., \& {West}, A.~A. 2011, \apj, 731, 17

\bibitem[{{Bell} {et~al.}(2012){Bell}, {Hilton}, {Davenport}, {Hawley}, {West},
  \& {Rogel}}]{bell2012}
{Bell}, K.~J., {Hilton}, E.~J., {Davenport}, J.~R.~A., {Hawley}, S.~L., {West},
  A.~A., \& {Rogel}, A.~B. 2012, PASP

\bibitem[{{Bochanski} {et~al.}(2010){Bochanski}, {Hawley}, {Covey}, {West},
  {Reid}, {Golimowski}, \& {Ivezi{\'c}}}]{bochanski2010}
{Bochanski}, J.~J., {Hawley}, S.~L., {Covey}, K.~R., {West}, A.~A., {Reid},
  I.~N., {Golimowski}, D.~A., \& {Ivezi{\'c}}, {\v Z}. 2010, \aj, 139, 2679

\bibitem[{{Bochanski} {et~al.}(2011){Bochanski}, {Hawley}, \&
  {West}}]{bochanski2011}
{Bochanski}, J.~J., {Hawley}, S.~L., \& {West}, A.~A. 2011, \aj, 141, 98

\bibitem[{{Bochanski} {et~al.}(2009){Bochanski}, {Hennawi}, {Simcoe},
  {Prochaska}, {West}, {Burgasser}, {Burles}, {Bernstein}, {Williams}, \&
  {Murphy}}]{bochanski2009}
{Bochanski}, J.~J., {et~al.} 2009, \pasp, 121, 1409

\bibitem[{{Bochanski} {et~al.}(2007{\natexlab{a}}){Bochanski}, {Munn},
  {Hawley}, {West}, {Covey}, \& {Schneider}}]{bochanski_gal}
{Bochanski}, J.~J., {Munn}, J.~A., {Hawley}, S.~L., {West}, A.~A., {Covey},
  K.~R., \& {Schneider}, D.~P. 2007{\natexlab{a}}, \aj, 134, 2418

\bibitem[{{Bochanski} {et~al.}(2007{\natexlab{b}}){Bochanski}, {West},
  {Hawley}, \& {Covey}}]{bochanski_templates}
{Bochanski}, J.~J., {West}, A.~A., {Hawley}, S.~L., \& {Covey}, K.~R.
  2007{\natexlab{b}}, \aj, 133, 531

\bibitem[{{Bonanos}(2009)}]{bonanos2009}
{Bonanos}, A.~Z. 2009, \apj, 691, 407

\bibitem[{{Borucki} {et~al.}(2010){Borucki}, {Koch}, {Basri}, {Batalha},
  {Brown}, {Caldwell}, {Caldwell}, {Christensen-Dalsgaard}, {Cochran},
  {DeVore}, {Dunham}, {Dupree}, {Gautier}, {Geary}, {Gilliland}, {Gould},
  {Howell}, {Jenkins}, {Kondo}, {Latham}, {Marcy}, {Meibom}, {Kjeldsen},
  {Lissauer}, {Monet}, {Morrison}, {Sasselov}, {Tarter}, {Boss}, {Brownlee},
  {Owen}, {Buzasi}, {Charbonneau}, {Doyle}, {Fortney}, {Ford}, {Holman},
  {Seager}, {Steffen}, {Welsh}, {Rowe}, {Anderson}, {Buchhave}, {Ciardi},
  {Walkowicz}, {Sherry}, {Horch}, {Isaacson}, {Everett}, {Fischer}, {Torres},
  {Johnson}, {Endl}, {MacQueen}, {Bryson}, {Dotson}, {Haas}, {Kolodziejczak},
  {Van Cleve}, {Chandrasekaran}, {Twicken}, {Quintana}, {Clarke}, {Allen},
  {Li}, {Wu}, {Tenenbaum}, {Verner}, {Bruhweiler}, {Barnes}, \&
  {Prsa}}]{borucki2010}
{Borucki}, W.~J., {et~al.} 2010, Science, Volume 327, Issue 5968, pp.~977
  (2010)., 327, 977

\bibitem[{{Bradstreet} \& {Guinan}(1994)}]{bradstreet1994}
{Bradstreet}, D.~H., \& {Guinan}, E.~F. 1994, in Astronomical Society of the
  Pacific Conference Series, Vol.~56, Interacting Binary Stars, ed. A.~W.
  {Shafter}, 228

\bibitem[{{Caballero-Garc{\'{\i}}a} {et~al.}(2010){Caballero-Garc{\'{\i}}a},
  {Torres}, {Ribas}, {R{\'{\i}}squez}, {Montesinos}, {Mas-Hesse}, \&
  {Domingo}}]{garcia2010}
{Caballero-Garc{\'{\i}}a}, M.~D., {Torres}, G., {Ribas}, I., {R{\'{\i}}squez},
  D., {Montesinos}, B., {Mas-Hesse}, J.~M., \& {Domingo}, A. 2010, \aap, 514,
  A36

\bibitem[{{Chabrier} {et~al.}(2007){Chabrier}, {Gallardo}, \&
  {Baraffe}}]{chabrier2007}
{Chabrier}, G., {Gallardo}, J., \& {Baraffe}, I. 2007, \aap, 472, L17

\bibitem[{{Chau}(1978)}]{chau1978}
{Chau}, W.~Y. 1978, \apj, 219, 1038

\bibitem[{{Covey} {et~al.}(2007){Covey}, {Ivezi{\'c}}, {Schlegel},
  {Finkbeiner}, {Padmanabhan}, {Lupton}, {Ag{\"u}eros}, {Bochanski}, {Hawley},
  {West}, {Seth}, {Kimball}, {Gogarten}, {Claire}, {Haggard}, {Kaib},
  {Schneider}, \& {Sesar}}]{covey2007}
{Covey}, K.~R., {et~al.} 2007, \aj, 134, 2398

\bibitem[{{Davenport} {et~al.}(2012){Davenport}, {Becker}, {Kowalski},
  {Hawley}, {Schmidt}, {Hilton}, {Sesar}, \& {Cutri}}]{davenport2012}
{Davenport}, J.~R.~A., {Becker}, A.~C., {Kowalski}, A.~F., {Hawley}, S.~L.,
  {Schmidt}, S.~J., {Hilton}, E.~J., {Sesar}, B., \& {Cutri}, R. 2012, \apj,
  748, 58

\bibitem[{{Dimitrov} \& {Kjurkchieva}(2010)}]{dimitrov2010}
{Dimitrov}, D.~P., \& {Kjurkchieva}, D.~P. 2010, \mnras, 406, 2559

\bibitem[{{Eggleton}(1983)}]{eggleton1983}
{Eggleton}, P.~P. 1983, \apj, 268, 368

\bibitem[{{Ford}(2006)}]{ford2006}
{Ford}, E.~B. 2006, \apj, 642, 505

\bibitem[{{Foreman-Mackey} {et~al.}(2012){Foreman-Mackey}, {Hogg}, {Lang}, \&
  {Goodman}}]{emcee}
{Foreman-Mackey}, D., {Hogg}, D.~W., {Lang}, D., \& {Goodman}, J. 2012, ArXiv
  e-prints

\bibitem[{{Friedman}(1984)}]{supersmoother}
{Friedman}, J.~H. 1984, A Variable Span Smoother, Tech. Rep.~5, Department of
  Statistics, Stanford University

\bibitem[{{Frieman} {et~al.}(2008){Frieman}, {Bassett}, {Becker}, {Choi},
  {Cinabro}, {DeJongh}, {Depoy}, {Dilday}, {Doi}, {Garnavich}, {Hogan},
  {Holtzman}, {Im}, {Jha}, {Kessler}, {Konishi}, {Lampeitl}, {Marriner},
  {Marshall}, {McGinnis}, {Miknaitis}, {Nichol}, {Prieto}, {Riess}, {Richmond},
  {Romani}, {Sako}, {Schneider}, {Smith}, {Takanashi}, {Tokita}, {van der
  Heyden}, {Yasuda}, {Zheng}, {Adelman-McCarthy}, {Annis}, {Assef},
  {Barentine}, {Bender}, {Blandford}, {Boroski}, {Bremer}, {Brewington},
  {Collins}, {Crotts}, {Dembicky}, {Eastman}, {Edge}, {Edmondson}, {Elson},
  {Eyler}, {Filippenko}, {Foley}, {Frank}, {Goobar}, {Gueth}, {Gunn},
  {Harvanek}, {Hopp}, {Ihara}, {Ivezi{\'c}}, {Kahn}, {Kaplan}, {Kent},
  {Ketzeback}, {Kleinman}, {Kollatschny}, {Kron}, {Krzesi{\'n}ski}, {Lamenti},
  {Leloudas}, {Lin}, {Long}, {Lucey}, {Lupton}, {Malanushenko}, {Malanushenko},
  {McMillan}, {Mendez}, {Morgan}, {Morokuma}, {Nitta}, {Ostman}, {Pan},
  {Rockosi}, {Romer}, {Ruiz-Lapuente}, {Saurage}, {Schlesinger}, {Snedden},
  {Sollerman}, {Stoughton}, {Stritzinger}, {Subba Rao}, {Tucker}, {Vaisanen},
  {Watson}, {Watters}, {Wheeler}, {Yanny}, \& {York}}]{frieman2008}
{Frieman}, J.~A., {et~al.} 2008, \aj, 135, 338

\bibitem[{{Grundstrom} {et~al.}(2007){Grundstrom}, {Gies}, {Hillwig},
  {McSwain}, {Smith}, {Gehrz}, {Stahl}, \& {Kaufer}}]{grundstrom2007}
{Grundstrom}, E.~D., {Gies}, D.~R., {Hillwig}, T.~C., {McSwain}, M.~V.,
  {Smith}, N., {Gehrz}, R.~D., {Stahl}, O., \& {Kaufer}, A. 2007, \apj, 667,
  505

\bibitem[{{Hilton} {et~al.}(2010){Hilton}, {West}, {Hawley}, \&
  {Kowalski}}]{hilton10}
{Hilton}, E.~J., {West}, A.~A., {Hawley}, S.~L., \& {Kowalski}, A.~F. 2010,
  \aj, 140, 1402

\bibitem[{{Holtzman} {et~al.}(2010){Holtzman}, {Harrison}, \&
  {Coughlin}}]{holtzman2010}
{Holtzman}, J.~A., {Harrison}, T.~E., \& {Coughlin}, J.~L. 2010, Advances in
  Astronomy, 2010

\bibitem[{{Ivezi{\'c}} {et~al.}(2007){Ivezi{\'c}}, {Smith}, {Miknaitis}, {Lin},
  {Tucker}, {Lupton}, {Gunn}, {Knapp}, {Strauss}, {Sesar}, {Doi}, {Tanaka},
  {Fukugita}, {Holtzman}, {Kent}, {Yanny}, {Schlegel}, {Finkbeiner},
  {Padmanabhan}, {Rockosi}, {Juri{\'c}}, {Bond}, {Lee}, {Stoughton}, {Jester},
  {Harris}, {Harding}, {Morrison}, {Brinkmann}, {Schneider}, \&
  {York}}]{ivezic2007}
{Ivezi{\'c}}, {\v Z}., {et~al.} 2007, \aj, 134, 973

\bibitem[{{Jiang} {et~al.}(2011){Jiang}, {Han}, {Ge}, {Yang}, \&
  {Li}}]{jiang2011}
{Jiang}, D., {Han}, Z., {Ge}, H., {Yang}, L., \& {Li}, L. 2011, ArXiv e-prints

\bibitem[{{Jiang} {et~al.}(2012){Jiang}, {Han}, {Ge}, {Yang}, \&
  {Li}}]{jiang2012}
---. 2012, \mnras, 421, 2769

\bibitem[{{Kopal}(1959)}]{kopal1959}
{Kopal}, Z. 1959, {Close binary systems}

\bibitem[{{Kowalski} {et~al.}(2009){Kowalski}, {Hawley}, {Hilton}, {Becker},
  {West}, {Bochanski}, \& {Sesar}}]{k09}
{Kowalski}, A.~F., {Hawley}, S.~L., {Hilton}, E.~J., {Becker}, A.~C., {West},
  A.~A., {Bochanski}, J.~J., \& {Sesar}, B. 2009, \aj, 138, 633

\bibitem[{{Kozai}(1962)}]{kozai1962}
{Kozai}, Y. 1962, \aj, 67, 591

\bibitem[{{Kruse} {et~al.}(2010){Kruse}, {Berger}, {Knapp}, {Laskar}, {Gunn},
  {Loomis}, {Lupton}, \& {Schlegel}}]{kruse2010}
{Kruse}, E.~A., {Berger}, E., {Knapp}, G.~R., {Laskar}, T., {Gunn}, J.~E.,
  {Loomis}, C.~P., {Lupton}, R.~H., \& {Schlegel}, D.~J. 2010, \apj, 722, 1352

\bibitem[{{Kubiak} {et~al.}(2006){Kubiak}, {Udalski}, \&
  {Szymanski}}]{kubiak2006}
{Kubiak}, M., {Udalski}, A., \& {Szymanski}, M.~K. 2006, ACTAA, 56, 253

\bibitem[{{Lada}(2006)}]{2006ApJ...640L..63L}
{Lada}, C.~J. 2006, \apjl, 640, L63

\bibitem[{{Li} {et~al.}(2004){Li}, {Han}, \& {Zhang}}]{li2004}
{Li}, L., {Han}, Z., \& {Zhang}, F. 2004, \mnras, 355, 1383

\bibitem[{{L{\'o}pez-Morales}(2007)}]{lopez2007}
{L{\'o}pez-Morales}, M. 2007, \apj, 660, 732

\bibitem[{{Maceroni} \& {Montalb{\'a}n}(2004)}]{maceroni2004}
{Maceroni}, C., \& {Montalb{\'a}n}, J. 2004, \aap, 426, 577

\bibitem[{{Maceroni} \& {Rucinski}(1997)}]{maceroni1997}
{Maceroni}, C., \& {Rucinski}, S.~M. 1997, \pasp, 109, 782

\bibitem[{{Mateo} {et~al.}(1990){Mateo}, {Harris}, {Nemec}, \&
  {Olszewski}}]{mateo1990}
{Mateo}, M., {Harris}, H.~C., {Nemec}, J., \& {Olszewski}, E.~W. 1990, \aj,
  100, 469

\bibitem[{{Maxted} {et~al.}(1998){Maxted}, {Marsh}, {Moran}, {Dhillon}, \&
  {Hilditch}}]{maxted1998}
{Maxted}, P.~F.~L., {Marsh}, T.~R., {Moran}, C., {Dhillon}, V.~S., \&
  {Hilditch}, R.~W. 1998, \mnras, 300, 1225

\bibitem[{{Mazeh}(2008)}]{mazeh2008}
{Mazeh}, T. 2008, in EAS Publications Series, Vol.~29, EAS Publications Series,
  ed. M.-J. {Goupil} \& J.-P. {Zahn}, 1--65

\bibitem[{{Morales} {et~al.}(2010){Morales}, {Gallardo}, {Ribas}, {Jordi},
  {Baraffe}, \& {Chabrier}}]{morales2010}
{Morales}, J.~C., {Gallardo}, J., {Ribas}, I., {Jordi}, C., {Baraffe}, I., \&
  {Chabrier}, G. 2010, \apj, 718, 502

\bibitem[{{Morgan} {et~al.}(2012){Morgan}, {West}, {Garc{\'e}s}, {Catal{\'a}n},
  {Dhital}, {Fuchs}, \& {Silvestri}}]{morgan2012}
{Morgan}, D.~P., {West}, A.~A., {Garc{\'e}s}, A., {Catal{\'a}n}, S., {Dhital},
  S., {Fuchs}, M., \& {Silvestri}, N.~M. 2012, ArXiv e-prints, 1205.6806

\bibitem[{{Nefs} {et~al.}(2012){Nefs}, {Birkby}, {Snellen}, {Hodgkin},
  {Pinfield}, {Sipocz}, {Kovacs}, {Mislis}, {Saglia}, {Koppenhofer}, {Cruz},
  {Barrado}, {Martin}, {Goulding}, {Stoev}, {Zendejas}, {del Burgo},
  {Cappetta}, \& {Pavlenko}}]{nefs2012}
{Nefs}, S.~V., {et~al.} 2012, ArXiv e-prints, 1206.1200

\bibitem[{{Pilecki} {et~al.}(2007){Pilecki}, {Fabrycky}, \&
  {Poleski}}]{pilecki2007}
{Pilecki}, B., {Fabrycky}, D., \& {Poleski}, R. 2007, \mnras, 378, 757

\bibitem[{{Pribulla} \& {Rucinski}(2008)}]{pribulla2008}
{Pribulla}, T., \& {Rucinski}, S.~M. 2008, \mnras, 386, 377

\bibitem[{{Pr{\v s}a} {et~al.}(2011){Pr{\v s}a}, {Batalha}, {Slawson}, {Doyle},
  {Welsh}, {Orosz}, {Seager}, {Rucker}, {Mjaseth}, {Engle}, {Conroy},
  {Jenkins}, {Caldwell}, {Koch}, \& {Borucki}}]{prsa2011}
{Pr{\v s}a}, A., {et~al.} 2011, \aj, 141, 83

\bibitem[{{Pr{\v s}a} \& {Zwitter}(2005)}]{phoebe}
{Pr{\v s}a}, A., \& {Zwitter}, T. 2005, \apj, 628, 426

\bibitem[{{Reid} \& {Hawley}(2005)}]{nlds2}
{Reid}, I.~N., \& {Hawley}, S.~L. 2005, {New light on dark stars : red dwarfs,
  low-mass stars, brown dwarfs}, ed. {Reid, I.~N.~\& Hawley, S.~L.}

\bibitem[{{Reimann}(1994)}]{reimann1994}
{Reimann}, J.~D. 1994, PhD thesis, UNIVERSITY OF CALIFORNIA, BERKELEY.

\bibitem[{{Ribas}(2006)}]{ribas2006}
{Ribas}, I. 2006, \apss, 304, 89

\bibitem[{{Rucinski}(1992)}]{rucinski1992}
{Rucinski}, S.~M. 1992, \aj, 103, 960

\bibitem[{{Rucinski}(1997)}]{ibvs4460}
---. 1997, Information Bulletin on Variable Stars, 4460, 1

\bibitem[{{Rucinski}(1998)}]{1998AJ....116.2998R}
---. 1998, \aj, 116, 2998

\bibitem[{{Rucinski}(2002)}]{rucinski2002}
---. 2002, \aj, 124, 1746

\bibitem[{{Rucinski} \& {Pribulla}(2008)}]{rucinski2008}
{Rucinski}, S.~M., \& {Pribulla}, T. 2008, \mnras, 388, 1831

\bibitem[{{Sesar} {et~al.}(2007){Sesar}, {Ivezi{\'c}}, {Lupton}, {Juri{\'c}},
  {Gunn}, {Knapp}, {DeLee}, {Smith}, {Miknaitis}, {Lin}, {Tucker}, {Doi},
  {Tanaka}, {Fukugita}, {Holtzman}, {Kent}, {Yanny}, {Schlegel}, {Finkbeiner},
  {Padmanabhan}, {Rockosi}, {Bond}, {Lee}, {Stoughton}, {Jester}, {Harris},
  {Harding}, {Brinkmann}, {Schneider}, {York}, {Richmond}, \& {Vanden
  Berk}}]{sesar2007}
{Sesar}, B., {et~al.} 2007, \aj, 134, 2236

\bibitem[{{Silvestri} {et~al.}(2006){Silvestri}, {Hawley}, {West}, {Szkody},
  {Bochanski}, {Eisenstein}, {McGehee}, {Schmidt}, {Smith}, {Wolfe}, {Harris},
  {Kleinman}, {Liebert}, {Nitta}, {Barentine}, {Brewington}, {Brinkmann},
  {Harvanek}, {Krzesi{\'n}ski}, {Long}, {Neilsen}, {Schneider}, \&
  {Snedden}}]{silvestri2006}
{Silvestri}, N.~M., {et~al.} 2006, \aj, 131, 1674

\bibitem[{{Stepien}(2006{\natexlab{a}})}]{stepien2006a}
{Stepien}, K. 2006{\natexlab{a}}, AcA, 56, 199

\bibitem[{{Stepien}(2006{\natexlab{b}})}]{stepien2006b}
---. 2006{\natexlab{b}}, AcA, 56, 347

\bibitem[{{Tegmark} {et~al.}(2004){Tegmark}, {Strauss}, {Blanton}, {Abazajian},
  {Dodelson}, {Sandvik}, {Wang}, {Weinberg}, {Zehavi}, {Bahcall}, {Hoyle},
  {Schlegel}, {Scoccimarro}, {Vogeley}, {Berlind}, {Budavari}, {Connolly},
  {Eisenstein}, {Finkbeiner}, {Frieman}, {Gunn}, {Hui}, {Jain}, {Johnston},
  {Kent}, {Lin}, {Nakajima}, {Nichol}, {Ostriker}, {Pope}, {Scranton},
  {Seljak}, {Sheth}, {Stebbins}, {Szalay}, {Szapudi}, {Xu}, {Annis},
  {Brinkmann}, {Burles}, {Castander}, {Csabai}, {Loveday}, {Doi}, {Fukugita},
  {Gillespie}, {Hennessy}, {Hogg}, {Ivezi{\'c}}, {Knapp}, {Lamb}, {Lee},
  {Lupton}, {McKay}, {Kunszt}, {Munn}, {O'Connell}, {Peoples}, {Pier},
  {Richmond}, {Rockosi}, {Schneider}, {Stoughton}, {Tucker}, {vanden Berk},
  {Yanny}, \& {York}}]{tegmark2004}
{Tegmark}, M., {et~al.} 2004, \prd, 69, 103501

\bibitem[{{Tylenda} {et~al.}(2011){Tylenda}, {Hajduk}, {Kami{\'n}ski},
  {Udalski}, {Soszy{\'n}ski}, {Szyma{\'n}ski}, {Kubiak}, {Pietrzy{\'n}ski},
  {Poleski}, {Wyrzykowski}, \& {Ulaczyk}}]{tylenda2011}
{Tylenda}, R., {et~al.} 2011, \aap, 528, A114

\bibitem[{{Udalski} {et~al.}(1992){Udalski}, {Szymanski}, {Kaluzny}, {Kubiak},
  \& {Mateo}}]{ogle}
{Udalski}, A., {Szymanski}, M., {Kaluzny}, J., {Kubiak}, M., \& {Mateo}, M.
  1992, ACTAA, 42, 253

\bibitem[{{Vanden Berk} {et~al.}(2004){Vanden Berk}, {Wilhite}, {Kron},
  {Anderson}, {Brunner}, {Hall}, {Ivezi{\'c}}, {Richards}, {Schneider}, {York},
  {Brinkmann}, {Lamb}, {Nichol}, \& {Schlegel}}]{vandenberk2004}
{Vanden Berk}, D.~E., {et~al.} 2004, \apj, 601, 692

\bibitem[{{West} {et~al.}(2008){West}, {Hawley}, {Bochanski}, {Covey}, {Reid},
  {Dhital}, {Hilton}, \& {Masuda}}]{west2008}
{West}, A.~A., {Hawley}, S.~L., {Bochanski}, J.~J., {Covey}, K.~R., {Reid},
  I.~N., {Dhital}, S., {Hilton}, E.~J., \& {Masuda}, M. 2008, \aj, 135, 785

\bibitem[{{West} {et~al.}(2011){West}, {Morgan}, {Bochanski}, {Andersen},
  {Bell}, {Kowalski}, {Davenport}, {Hawley}, {Schmidt}, {Bernat}, {Hilton},
  {Muirhead}, {Covey}, {Rojas-Ayala}, {Schlawin}, {Gooding}, {Schluns},
  {Dhital}, {Pineda}, \& {Jones}}]{westdr7}
{West}, A.~A., {et~al.} 2011, \aj, 141, 97

\bibitem[{{Wilson}(2001)}]{ibvs5076}
{Wilson}, R.~E. 2001, Information Bulletin on Variable Stars, 5076, 1

\bibitem[{{Wilson} \& {Devinney}(1971)}]{wilson1971}
{Wilson}, R.~E., \& {Devinney}, E.~J. 1971, \apj, 166, 605

\bibitem[{{York} {et~al.}(2000){York}, {Adelman}, {Anderson}, {Anderson},
  {Annis}, {Bahcall}, {Bakken}, {Barkhouser}, {Bastian}, {Berman}, {Boroski},
  {Bracker}, {Briegel}, {Briggs}, {Brinkmann}, {Brunner}, {Burles}, {Carey},
  {Carr}, {Castander}, {Chen}, {Colestock}, {Connolly}, {Crocker}, {Csabai},
  {Czarapata}, {Davis}, {Doi}, {Dombeck}, {Eisenstein}, {Ellman}, {Elms},
  {Evans}, {Fan}, {Federwitz}, {Fiscelli}, {Friedman}, {Frieman}, {Fukugita},
  {Gillespie}, {Gunn}, {Gurbani}, {de Haas}, {Haldeman}, {Harris}, {Hayes},
  {Heckman}, {Hennessy}, {Hindsley}, {Holm}, {Holmgren}, {Huang}, {Hull},
  {Husby}, {Ichikawa}, {Ichikawa}, {Ivezi{\'c}}, {Kent}, {Kim}, {Kinney},
  {Klaene}, {Kleinman}, {Kleinman}, {Knapp}, {Korienek}, {Kron}, {Kunszt},
  {Lamb}, {Lee}, {Leger}, {Limmongkol}, {Lindenmeyer}, {Long}, {Loomis},
  {Loveday}, {Lucinio}, {Lupton}, {MacKinnon}, {Mannery}, {Mantsch}, {Margon},
  {McGehee}, {McKay}, {Meiksin}, {Merelli}, {Monet}, {Munn}, {Narayanan},
  {Nash}, {Neilsen}, {Neswold}, {Newberg}, {Nichol}, {Nicinski}, {Nonino},
  {Okada}, {Okamura}, {Ostriker}, {Owen}, {Pauls}, {Peoples}, {Peterson},
  {Petravick}, {Pier}, {Pope}, {Pordes}, {Prosapio}, {Rechenmacher}, {Quinn},
  {Richards}, {Richmond}, {Rivetta}, {Rockosi}, {Ruthmansdorfer}, {Sandford},
  {Schlegel}, {Schneider}, {Sekiguchi}, {Sergey}, {Shimasaku}, {Siegmund},
  {Smee}, {Smith}, {Snedden}, {Stone}, {Stoughton}, {Strauss}, {Stubbs},
  {SubbaRao}, {Szalay}, {Szapudi}, {Szokoly}, {Thakar}, {Tremonti}, {Tucker},
  {Uomoto}, {Vanden Berk}, {Vogeley}, {Waddell}, {Wang}, {Watanabe},
  {Weinberg}, {Yanny}, \& {Yasuda}}]{york2000}
{York}, D.~G., {et~al.} 2000, \aj, 120, 1579

\bibitem[{{Zucker} \& {Mazeh}(1994)}]{todcor}
{Zucker}, S., \& {Mazeh}, T. 1994, \apj, 420, 806

\end{thebibliography}

\end{document}